# In silico investigation of lactone and thiolactone inhibitors in bacterial quorum sensing using molecular modeling

--Manuscript Draft--

| | |
|---|---|
| **Manuscript Number:** | TCAC-D-13-00101 |
| **Full Title:** | In silico investigation of lactone and thiolactone inhibitors in bacterial quorum sensing using molecular modeling |
| **Article Type:** | Regular Article |
| **Keywords:** | Quorum Sensing; lactone inhibitors; Docking; Molecular dynamics; AMBER |
| **Corresponding Author:** | Marawan M Ahmed, Bsc<br>Swinburne University of technology<br>Melbourne, Non-US/Canada AUSTRALIA |
| **Corresponding Author Secondary Information:** | |
| **Corresponding Author's Institution:** | Swinburne University of technology |
| **Corresponding Author's Secondary Institution:** | |
| **First Author:** | Marawan M Ahmed, Bsc |
| **First Author Secondary Information:** | |
| **Order of Authors:** | Marawan M Ahmed, Bsc |
| | Stefanie Bird, Student |
| | Feng Wang, Professor |
| | Enzo A. Palombo, Professor |
| **Order of Authors Secondary Information:** | |
| **Abstract:** | In the present study, the origin of the anti-quorum sensing (QS) activities of several members of a recently synthesized and in vitro tested class of lactone and thiolactone based inhibitors were computationally investigated. Docking and molecular dynamic (MD) simulations and binding free energy calculations were carried out to reveal the exact binding and inhibitory profiles of these compounds. The higher in vitro activity of the lactone series relative to their thiolactone isosteres was verified based on estimating the binding energies, the docking scores and monitoring the stability of the complexes produced in the MD simulations. The strong electrostatic contribution to the binding energies may be responsible for the higher inhibitory activity of the lactone with respect to the thiolactone series. The results of this study help to understand the anti-QS properties of lactone-based inhibitors and provide important information that may assist in the synthesis of novel QS inhibitors. |
| **Suggested Reviewers:** | Chang-Guo Zhan, Professor<br>College of Pharmacy, University of Kentuc<br>zhan@uky.edu |
| | Jaime Rubio-Martinez, Professor<br>Universitat de Barcelona (UB) and the Institut de Recerca en Química Teòrica i Computacional (IQTCUB), Martí i Franqués 1, 08028, Barcelona, Spain<br>jaime.rubio@ub.edu |
| | Prasad V. Bharatam, Professor<br>National Institute of Pharmaceutical Education and Research (NIPER), Sector-67, S. A. S. Nagar, 160 062, Punjab, India<br>pvbharatam@niper.ac.in |
| | Dawei Zhang<br>School of Physical and Mathematical Sciences, Nanyang Technological University, Singapore, 637371, Singapore |
| | zhangdw@ntu.edu.sg |
| | Modesto Orozco<br>Institut de Recerca Biomedica Parc Cientific de Barcelona Barcelona, Spain<br>modesto@mmb.pcb.ub.es |

# *In silico* investigation of lactone and thiolactone inhibitors in bacterial quorum sensing using molecular modeling

Marawan Ahmed[a,b]*, Stefanie Bird[a], Feng Wang[a,b]* and Enzo A. Palombo[b]

[a]eChemistry Laboratory, Faculty of Life and Social Sciences, Swinburne University, P.O. Box 218, Hawthorn, Victoria 3122, Australia.

[b] Environment and Biotechnology Centre, Faculty of Life and Social Sciences, Swinburne University, P.O. Box 218, Hawthorn, Victoria 3122, Australia.

## Abstract

In the present study, the origin of the anti-quorum sensing (QS) activities of several members of a recently synthesized and *in vitro* tested class of lactone and thiolactone based inhibitors were computationally investigated. Docking and molecular dynamic (MD) simulations and binding free energy calculations were carried out to reveal the exact binding and inhibitory profiles of these compounds. The higher *in vitro* activity of the lactone series relative to their thiolactone isosteres was verified based on estimating the binding energies, the docking scores and monitoring the stability of the complexes produced in the MD simulations. The strong electrostatic contribution to the binding energies may be responsible for the higher inhibitory activity of the lactone with respect to the thiolactone series. The results of this study help to understand the anti-QS properties of lactone-based inhibitors and provide important information that may assist in the synthesis of novel QS inhibitors.

* Correspondence authors: mmahmed@swin.edu.au (Tel: 61-3-9214 8785) and fwang@swin.edu.au (Tel. 61-3-9214-5065).

## 1. Introduction

Treatment of bacterial infections is a major global challenge. Bacteria continue to develop resistance to current anti-bacterial agents and the problem is becoming more wide-spread [1-4]. It is estimated that bacterial resistance can increase mortality and morbidity by a factor of two [5]. The problem is even worse in developing countries where appropriate medical services cannot always be effectively delivered [6]. An attractive pathway to resolve the problem of resistance is targeting bacterial quorum sensing [7-11].

Quorum sensing (QS) is a communication mechanism by which bacterial cells organize biological processes that are not possible with a single bacterium, such as toxin production and biofilm formation [8,12,13]. This mechanism includes binding of specific signal "hormone-like" molecules called "auto-inducers" to specific intracellular or membrane bound receptors [8,12,14]. This binding triggers a wide range of intracellular reaction cascades in Gram positive and Gram negative bacteria to carry out the required biological process [15,14]. A typical QS system is composed of three components, (i) A bacterial synthase (e.g. LuxI) that synthesizes the "auto-inducer", (ii) The auto-inducer which is typically an acylhomoserine lactone (AHL) derivative and (iii) A transcription regulatory protein, such as LuxR or its homologues. LuxR protein binds to DNA and activates gene expression once the level of the AHL reaches a critical threshold depending on the bacterial population density [16,17].

Various mechanisms of this AHL induced QS activation have been proposed and three mechanisms are the most widely accepted. In the first mechanism, the AHL induced conformational changes on LuxR enable LuxR to bind DNA and trigger the transcription process [18,17]. In the second mechanism, the AHL induced LuxR conformational changes relieve the repressor effect exerted by LuxR on the target genes and enable gene transcription [19,20]. In the third mechanism, extracellular AHL is detected by membrane bound receptors that trigger a wide range of intracellular reactions leading to gene expression [21,8].

Bacteria cannot easily develop an acquired resistance against QS inhibitors. As a result, QS inhibition is seen as an excellent weapon to fight against bacteria [8,16]. A number of distinct methods have been described to inhibit QS. In one of such methods, the AHL synthase is inhibited by small molecule analogues of organic compounds involved in AHL biosynthesis [22-24]. Enzymatic hydrolysis of the AHL molecule by acylases, hydrolases and lactonases has been reported as an excellent defense mechanism for other organisms against bacteria [22-24]. The third and most widely investigated method is the use of small AHL analogues that competitively inhibit AHL binding to the LuxR proteins and their homologues, such as the LasR protein [24,9,7,25-28]. These classes of inhibitors are referred to as "AHL antagonists". In the present study, two *in vitro* tested lactone and thiolactone series of AHL antagonists are computationally investigated to understand the origin of their anti-QS activities.

The most commonly investigated AHL antagonists are those belonging to lactone, thiolactone and furanone classes of organic compounds [24,9,7,25-29]. This is due to the structural similarities between these molecules and the naturally occurring AHL auto-inducers. These auto-inducers can have different structures depending on the producing organism. In most cases, the auto-inducer is composed of a five-membered lactone head and an acyl group spacer connecting this head to a hydrophobic chain tail. The length of the tail and its chemical structure differ between AHL derivatives and can affect the potency and the intrinsic activity (agonist or antagonist) of a given AHL [27,22,30,8]. This may facilitate the tuning of the effect for a given inhibitor or inducer so that it can selectively inhibit or activate a particular type of bacterium. Some authors suggested the ability of several molecules to inhibit QS although they are not structurally related to AHL [31].

Unfortunately, computational studies on this important class of inhibitors are rare which may be due to the limited availability of crystal structures of the LuxR proteins complexed with their corresponding antagonists [28,32-34]. The need for detailed investigation of QS inhibition at the molecular level is necessary for the understanding of QS process and the future development of effective drugs. In the current study, the binding mode of a recently synthesized and tested thiolactone group of AHL antagonists against LuxR proteins [8] was investigated. Docking and molecular dynamics (MD) simulations were carried out against a recently resolved X-ray crystal structure of LuxR protein from *Chromobacterium violaceum (C. violaceeum)*. The *C. violaceum* LuxR protein (CviR) was co-crystallized with an inhibitor from a similar study but having a lactone ring instead of thiolactone, i.e., sulphur has been replaced by its isosteric oxygen atom [7]. In addition to the reported thiolactone series, the corresponding lactone analogues are computationally investigated to understand the differences in binding between the two groups. Also, it has been shown that the thiolactone analogue of the co-crystallized inhibitor is 10-fold less effective than the lactone inhibitor, a detailed analysis is carried out to understand the basis of this difference.

## 2. METHODS AND COMPUTATIONAL DETAILS

The chemical structure of the lactone and thiolactone back bone skeleton is given in Figure 1. When the X atom in the penton ring is oxygen, i.e. X=O, the structures are lactones, whereas the structures become thiolactones if X=S in the same figure. The R-group at the end of the chain in the back bone structure is replaced by different sixteen groups as listed in the figure, which in total produced 32 compounds: 16 thiolactones and 16 lactones. The thiolactone series is denoted as "**TL**", the lactone series is denoted as "**L**" as marked in the same figure, in which the original co-crystallized lactone inhibitor is denoted as **L3** and its thiolactone isostere is denoted as **TL3**.

In the protein preparation, the crystal structure of CviR (a LuxR protein) co-crystallized with chlorolactone (**L3**) antagonist was taken from the PDB (PDB entry: 3QP5) [7]. Figure 2 gives a three-dimensional (3D) representation of the CviR protein monomer. The complex was prepared using the protein preparation wizard in Maestro 9.2 [35]. The crystallized protein structure is a tetramer, one chain is kept, and others are deleted in the present study and saturated by hydrogen atoms but the water molecules are deleted. Similar to a precious study[36,37], the missing residues were added and refined using Prime 3.0 [38]. The N-acetyl (ACE) and N-methyl amide (NMA) groups were added to cap the uncapped N and C termini respectively. H-bond network optimization was carried out assuming a neutral pH of the solution. The protonation states of titratable amino acids were assigned at the same pH. An all atom impref minimization step was carried out to remove unfavorable steric clashes until a convergence was reached or with a maximum RMSD of 0.3 Å from the original conformation. No steric clashes were reported after the final minimization step.

Once the protein structure is set up, a receptor grid was prepared with the receptor grid generation module in Glide 5.8 [39]. The binding site was determined as a box around the ligand that was centered inside the box. Four H-bonds constraints with the nearby residues (Tyr80, Trp84 Asp97, and Ser155) were set in the grid preparation.

Ligand molecules were optimized at the RM1 [40] semiempirical level of theory as implemented in the Semiempirical module in Maestro 9.2 [35]. Ligand partial atomic electrostatic potential charges (ESP) charges were assigned at the HF/cc-pVTZ level of theory using Jaguar [41].

Next, docking and scoring of the study employed the flexible ligand docking, which was performed through the Glide extra precision mode (Glide XP) [42]. In order to increase the sampling space, a maximum of 50.000 initial ligand poses were kept in the initial phase of docking. A scoring window of poses within 1000 kcal·mol$^{-1}$ from the best scoring pose were retained, from which a maximum of 800 poses per ligand were subjected to 200 steps of energy minimization. A potential ligand pose was considered only when at least three of the four predetermined H-bond constraints were satisfied. Rescoring the docked poses was done using the Prime/MM-GBSA module in Prime 3.0; residues within 6Å of the ligand were considered flexible.

Finally, molecular dynamics (MD) simulations were conducted for the co-crystallized lactone inhibitors and its thiolactone analogues with both the dimeric and the monomeric forms of the CviR protein, i.e. four inhibitor-protein complexes were simulated. That is, the **L3**/**TL3**-CviR monomer complexes and the **L3**/**TL3**-CviR dimer complexes. To remove any potential bias from different starting configurations, the **TL3** complexes were obtained by mutating the oxygen atom of the experimentally resolved **L3** complexes to a sulphur atom.

The structure preparation and the following MD simulations were performed using AMBER 12 software package [43] applying the ff03 force field [44]. Single point calculations of the corresponding inhibitors were performed at the HF/6-31G* using the Gaussian 09 program [45]. The inhibitor charges and other parameters were obtained using the RESP fitting [46] procedures and the general AMBER force field (GAFF) [47]. The complexes were then solvated in a box of TIP3P [48] water with a buffer size of 15Å and were neutralized by counter ions.

Each system was then subjected to four consecutive minimization steps. In each step, water molecules and ions were allowed to move freely for a 1000 steps of steepest descent minimization followed by 4000 steps of conjugate gradient minimization holding protein and inhibitor atoms constrained to their original positions by a force constant of 100 $kcal \cdot mol^{-1} Å^{-2}$, then gradually releasing the force constraints to 50, 5 and zero (no constraints) $kcal \cdot mol^{-1} Å^{-2}$, respectively. Following minimization, two consecutive steps of heating and equilibration were performed. Each system was gradually heated in the NVT ensemble from 0°K to 300°K for 30 ps with a time step of 1 fs, applying a force constant of 10 $kcal \cdot mol^{-1} Å^{-2}$ on the protein and inhibitor coordinates. Langevin dynamics with the collision frequency $\gamma$ of 1 $ps^{-1}$ for temperature control was employed. A further 1 ns simulation in the NPT ensemble was performed to equilibrate the system density by applying a time step of 2 fs, which required the use of SHAKE algorithm [49] to constrain all bonds involving hydrogen atoms. The temperature was controlled using Langevin dynamics with the collision frequency $\gamma$ of 1 $ps^{-1}$ and is kept at 300°K. The pressure was kept at 1 bar by applying a Berendsen barostate with a pressure relaxation time of 1 ps. Each system was again relaxed in the NVT ensemble for 20 ns followed by 30 ns production simulation at 300°K using Berendsen temperature control [50]. In all simulation steps, long-range electrostatics were computed using the particle mesh Ewald (PME) and a 12 Å real space cut-off [51]. The edge effect was removed by applying periodic boundary conditions. For MD simulations of the monomeric chains, weak constraining forces were applied on the DNA binding domain plus the flexible coil segments of the monomeric chain. All MD simulations were carried out using the PMEMD module of AMBER12. For the binding energy evaluation from the trajectory, the MM/PBSA module of AMBER12 was used and using every second frame collected from the MD simulations, i.e. around 3750 snapshots were used [52].

## 3. RESULTS AND DISCUSSION

### 3.1. Docking and scoring

To the best of our knowledge, the first available X-ray crystal structure for a member of the LuxR family of proteins co-crystallized with a pure antagonist is the CviR protein from *Chromobacterium violaceum* (PDB code: 3QP5) [7]. In that study, CviR was co-crystallized with various ligands of either agonistic or antagonistic activities. Agonist binding to CviR results in conformational changes and activation of the dimer to bind DNA and trigger DNA transcription. The ligand induced conformational changes determine the intrinsic activity of a given ligand to be either an agonist or an antagonist. In addition, it has been shown that subtle ligand structural differences can affect the potency and the intrinsic activity of a given ligand dramatically [27,22,30,8]. Moreover, the same ligand can work as an agonist or an antagonist depending on the protein homologue and the bacterial strain [27,22,30,8].

These unique properties urge further and extensive theoretical and experimental work to shed the light on the complex mechanism that controls QS signaling.

Figure 2 presents a solid ribbon representation of a CviR protein monomer. The protein is made of two distinct domains, a Ligand Binding Domain (LBD) and a DNA Binding Domain (DBD) and the two domains are connected by a short flexible coil. The LBD is the bigger domain and is composed of α-helices and β-sheets while the DBD is composed of a few α-helices. The exact ligand binding site is shown as a solid surface inside the LBD.

AHLs or their analogues are characterized by a unique "sperm-like" structure composed of two parts, head and tail. The lactone head is able to form an H-bond with the nearby Trp84 residue, while the acyl group forms H-bonds with Asp97, Tyr80 and Ser155. The tail part is buried in a hydrophobic pocket made of Val, Leu and Ile residues. Figure 3 shows two 2D ligand interaction diagrams for an agonist (PDB code: 3QP1) and an antagonist (PDB code: 3QP5).

Table 1 reports the docking scores for the inhibitors under study using the conventional docking scores and the Prime-MM/GBSA scores. Inhibitors are given in the table according to their XP Gscore. From the correlation with the available experimental $IC_{50}$ data, the Prime MMGBSA DG bind vdW score achieved the best correlation with experimentally measured $IC_{50}$ ($r_{(pearson)}$ = 0.52). Glide Emodel performed reasonably well ($r_{(pearson)}$ = 0.49).The thiolactone derivatives showed overall lower scores than their corresponding lactone analogues, which is consistent with the experimental findings. Only the XP Gscore was able to identify the original lactone inhibitor (**L3**) followed by the **TL12** thiolactone derivative to be the best inhibitors among the two series.

Figure 4 represents the 2D and 3D interaction diagrams of some selected antagonists with the CviR receptor. In general, antagonists binding modes to the receptor are similar to the original co-crystallized antagonist. This is a result of the strict docking criteria which are applied for accepting poses. The lactone carbonyl forms a direct H-bond with the conserved Trp84 residue, the acyl group –NH forms an H-bond with Asp97 and the carbonyl oxygen forms H-bonds with Tyr80 and Ser155. As the libraries are focused, subtle ligand differences which are correlated directly with inhibitory activity need to be paid greater attention.

The main difference between the two libraries is the isosteric replacement of the "**S**" atom in the **TL** series by an "**O**" atom in the **L** series. The observed activity differences between the two libraries may be related to the H-bond strength that may exist between the sulphur or oxygen ring atoms and the nearby $–C_{(7)}H$ of Trp84. However, it is known that heterocyclic H-bond acceptors are grouped in the "weak H-bond" category of acceptors [53]. Therefore, potential large effects of this H-bond (if any) are not expected. The effect of this substitution is discussed in detail in section 3.2.

The major difference between any pair of inhibitors within the same library (**TL** or **L**) is the chemical structure of the hydrophobic tail chain. Studies have indicated that the structure and the length of this chain can affect the potency and the intrinsic activity of a given ligand [27,22,30,8]. In the two inhibitor libraries, the terminal aromatic group forms a direct, sandwich type π-π stacking interaction with Tyr88 aromatic ring. Assuming all inhibitors have a similar binding mode, the overall binding strength (within a given library) is directly correlated with the strength of this π-π stacking interaction.

As can be seen in Table 1, substitution with Electron Donating Groups (**EDG**), such as methoxy groups, resulted in a dramatic reduction in the activity regardless of the position of this substitution. They also had the overall worst docking scores. For example, the $IC_{50}$ of –meta (**TL10**) and –para (**TL7**) substituted methoxy group derivatives are 37 µM and 11 µM, respectively. On the other hand, Electron Withdrawing Group (**EWG**) derivatives, such as halogenated derivatives, have the highest inhibitory effect and best docking scores as well. For example, the top scoring inhibitors (according to the Glide XP score) were the **L3** and the **TL12** inhibitors, which have IC50 values of 0.38 µM and 0.63 µM, respectively. The lactone derivative of **TL12**, the **L12** antagonist, exhibited a higher XP Gscore (-10.02). This makes the halogenated derivatives, particularly the poly-halogenated ones, better candidates for further synthesis and biological testing.

### 3.2. Molecular dynamic simulations

A difficult challenge in performing a meaningful MD simulation for LuxR proteins (including CviR) is their inherent flexibility. This inherent flexibility is obvious knowing that the protein can adopt different conformations depending on their activation states as well as the accompanying ligand. This flexibility enables different proteins to carry out their functions properly [54]. CviR is a homo-dimer composed of two identical and overlapping chains of about 250 amino acids each. Each monomeric chain is composed of an LBD connected to a DBD through a highly flexible coil.

An accurate measure to assess the stability of the protein complexes during MD simulations is the root mean square deviations (RMSD). Figure 5 (a-b) displays the RMSD plots for the Cα atoms of each protein in the production simulation period. As can be seen in the plots, the **L3** complexes, in both the monomeric and dimeric forms, have higher stabilities than their **TL3** counterparts. The average RMSD for **L3**-CviR dimer is ~ 1.6-1.8 Å. For the **TL3**-CviR dimer complex, the average RMSD value is higher with an average value of ~4 Å. This illustrates the enhanced stability of the CviR dimer with **L3** over **TL3**. Interestingly, the same is true for the monomeric case such that the **L3** complex with the monomeric CviR is more stable than its **TL3** counterpart.

To examine the origin of this reduced stability of the **TL3** complexes with respect to the **L3** complexes, a more detailed analysis on a per-residue basis was conducted using the per-residue heavy atoms

RMSFs (root mean square fluctuations). Figure 6(a-b) displays the heavy atoms RMSF plots of the four complexes during the simulation time. As can be seen, the **L3** complexes show enhanced stability over the **TL3** complexes. For the CviR dimer complexes, the LBDs of both chains (A and B) possess an overall lower RMSF during the simulation period than the DBDs. For the monomeric complexes with **L3** and **TL3**, and as a result of the weak constraints applied on the "DBD+coil" segment of the monomeric chains, the whole monomer is of a comparable RMSF value. The enhanced flexibility of the DBD segment of the dimeric protein is expected given the fact that this segment is responsible for binding to DNA upon activation [55]. It would be also interesting to investigate the detailed binding event of this segment to DNA upon agonists or antagonists binding. However, such a complex process is beyond the capability of conventional MD simulation and other MD paradigms, such as accelerated MD, may be more suitable [56]. Research in this direction is currently in progress.

It is important to study the stability of the observed H-bonds as a function of the simulation time. Thus, Figure 7 (a-d) displays some selected H-bonds monitored during the simulation time for the four complexes. The first and most important H-bond is the one formed with Trp84 residues and it has been already identified in the docking section. As can be seen in Figure 7, the average value for this H-bond is ~ 1.9-2.3 Å for all complexes. The **L3** complexes have less fluctuation during the simulation time than their **TL3** counterparts. The same is true for the H-bonds formed with the Tyr80 residue in all complexes. For the **L3** and **TL3** complexes with the monomeric CviR chains, higher fluctuations of up to 3.5 Å (**L3**) and 4.3 Å (**TL3**) were observed. Interestingly, and in most cases, the H-bond which is observed via docking between the amide carbonyl and the Ser155 residue is not stable in the MD simulation and the adopted rotomer shifts the –OH group to the other side (Figure 8).

### 3.3. Total and decomposed MM-PB/GBSA binding energies

Table 2 reports the binding energy scores for **L3** and **TL3** inhibitors according to the AMBER-MMPB/GBSA scores as implemented in AMBER12. The AMBER-MMPB/GBSA binding energy scores take the advantage of statistical averaging over many potential conformations that are produced from the MD trajectories. In the AMBER-MMPB/GBSA calculations, only the inhibitor-dimer complexes are considered. To further enhance statistical precision and knowing that each complex contains two inhibitors, each inhibitor is treated separately as a ligand in a separate run. Final data reported are the average values of the two independent AMBER-MMPB/GBSA calculations for each complex.

As can be seen in Table 2, the major term that favors the binding for both inhibitors is the vdW lipophilic term ($\Delta E_{vdW}$). Interestingly, although the vast majority of the binding site residues are lipophilic residues, the electrostatic ($\Delta E_{ele}$) term still exhibits a significant contribution to the binding. This large contribution emphasizes the importance of the H-bond interactions which are electrostatic in nature. It would be also interesting to use certain advanced techniques, such as alanine-scanning [57,58], to study the effect of mutations in residues responsible for these H-bonding interactions. Research in this direction is currently in progress.

For **TL3**, the $\Delta E_{vdW}$ interaction is given by -46.56 kcal·mol$^{-1}$, this value is slightly higher than that for **L3** which is given by -45.31 kcal·mol$^{-1}$. On the other hand, **L3** exhibits a larger contribution from the $\Delta E_{ele}$ term (-33.54 kcal·mol$^{-1}$) than **TL3** (-28.37) i.e., $\Delta\Delta E_{ele}$ is equal to 5.17 kcal·mol$^{-1}$ which is almost equal to the energy contribution of a full H-bond. This larger contribution of the $\Delta E_{ele}$ term for **L3** than **TL3** may be responsible for the fact that **L3** is ~10 times more active than **TL3** in the *in vitro* assay [8,7]. Regarding total binding energies as expressed by the $\Delta G$ values, **L3** shows higher binding energy according to the AMBER-MM/PBSA score (-63.54 kcal·mol$^{-1}$) than **TL3** (-59.11 kcal·mol$^{-1}$). The same is true for the AMBER-MM/GBSA score of **L3** (-48.65 kcal·mol$^{-1}$) compared to that for **TL3** (-47.60 kcal·mol$^{-1}$).

Decomposition of the binding energy on a per-residue basis is very important to understand the binding mode and assess the role of each residue in the binding. Figure 9 displays the per-residue binding energy analyses for the two complexes during the simulation period. For the $\Delta E_{vdW}$ and the $\Delta E_{ele}$ interaction terms, only residues showing large contributions are selected. Figure 9a displays the per-residue contribution for the $\Delta E_{vdW}$ interaction term. As can be seen, the major contribution to the $\Delta E_{vdw}$ term is from the Tyr88 via a strong $\pi$-$\pi$ stacking interactions, consistent with the observations from the docking study. The contribution of this residue to the $\Delta E_{vdw}$ term for both inhibitors is similar, -2.57 kcal·mol$^{-1}$ for **L3** and -2.59 kcal·mol$^{-1}$ for **TL3**. This is because the terminal aromatic ring which is responsible for the $\pi$-$\pi$ stacking interaction with this residue is identical in both inhibitors (**L3** and **TL3**). The second most important $\Delta E_{vdW}$ interaction is from the Trp111 residue which contributes more to **TL3** (-2.13 kcal·mol$^{-1}$) than to **L3** (-1.96 kcal·mol$^{-1}$) as a result of the presence of the more lipophilic sulphur in **TL3** instead of oxygen in **L3**.

Figure 9b displays the per-residue contribution for the $\Delta E_{ele}$ interaction term. The most important residue is Asp97 which contributes almost equally for the two inhibitors (-11.83 kcal·mol$^{-1}$ for **L3** and -11.35 kcal·mol$^{-1}$ for **TL3**). Differentiation for the $\Delta E_{ele}$ term between the two inhibitors is from the Tyr80 and Trp84 residues which show the highest discrepancy between the two inhibitors. For **L3**, the $\Delta E_{ele}$ contributions from the Tyr80 and the Trp84 residues are -3.37 kcal·mol$^{-1}$ and -2.88 kcal·mol$^{-1}$, respectively. For **TL3**, the contributions of these two residues are -1.87 kcal·mol$^{-1}$ and -2.03 kcal·mol$^{-1}$, respectively. The $\Delta\Delta E_{ele}$ contribution between the two inhibitors from these two residues together is 2.35 kcal·mol$^{-1}$ which is almost half of the energy contribution of a full H-bond. This emphasizes the importance of these two residues for any future development of CviR antagonists.

## 4. CONCLUSIONS

The exact binding modes of a series of recently synthesized and *in vitro* tested potential QS inhibitors were investigated *in silico*. Consistent with the experimentally measured $IC_{50}$ values, this molecular modeling study using docking scores and energies of binding showed that the lactone based inhibitors indeed exhibit stronger binding properties than the thiolactone based class of inhibitors. Molecular dynamics simulations of different inhibitor-protein complexes further indicated that the lactone based inhibitors were more stable than the thiolactone based ones. Relative to the number of hydrophilic residues present in the binding site, the electrostatic effect made a significant contribution to the binding for the two series of inhibitors. Certain residues, such as Tyr80 and Trp84, were discriminating between the two series of inhibitors with a larger electrostatic contribution for the lactone than the thiolactone inhibitors.

## ACKNOWLEDGEMENTS

MA acknowledges the Swinburne University Postgraduate Research Award (SUPRA). FW thanks the National Computational Infrastructure (NCI) at the Australian National University under the Merit Allocation Scheme (MAS), the Victorian Partnership for Advanced Computing (VPAC), Swinburne University supercomputing (Green/gSTAR) and the Victorian Life Sciences Computation Initiative (VLSCI) facilities.

## References

1. Levy SB, Marshall B (2004) Antibacterial resistance worldwide: causes, challenges and responses. Nat Med 10:122-129.
2. Levy SB (2001) Antibiotic Resistance: Consequences of Inaction. Clin Infect Dis 33:124-129.
3. Levy SB (2002) Factors impacting on the problem of antibiotic resistance. J Antimicrob Chemother 49:25-30.
4. Aiello AE, Larson E (2003) Antibacterial cleaning and hygiene products as an emerging risk factor for antibiotic resistance in the community. The Lancet infectious diseases 3:501-506.
5. Holmberg SD, Solomon SL, Blake PA (1987) Health and economic-impacts of antimicrobial resistance Reviews of Infectious Diseases 9:1065-1078.
6. Okeke IN, Laxminarayan R, Bhutta ZA, Duse AG, Jenkins P, O'Brien TF, Pablos-Mendez A, Klugman KP (2005) Antimicrobial resistance in developing countries. Part I: recent trends and current status. The Lancet infectious diseases 5:481-493.
7. Chen G, Swem Lee R, Swem Danielle L, Stauff Devin L, O'Loughlin Colleen T, Jeffrey Philip D, Bassler Bonnie L, Hughson Frederick M (2011) A Strategy for Antagonizing Quorum Sensing. Mol Cell 42:199-209.
8. Swem LR, Swem DL, O'Loughlin CT, Gatmaitan R, Zhao B, Ulrich SM, Bassler BL (2009) A quorum-sensing antagonist targets both membrane-bound and cytoplasmic receptors and controls bacterial pathogenicity. Mol Cell 35:143-153.
9. Galloway WRJD, Hodgkinson JT, Bowden SD, Welch M, Spring DR (2011) Quorum sensing in Gram-negative bacteria: Small-molecule modulation of AHL and AI-2 quorum sensing pathways. Chem Rev 111:28.
10. Suga H, Smith KM (2003) Molecular mechanisms of bacterial quorum sensing as a new drug target. Curr Opin Chem Biol 7:586-591.
11. Kalia VC, Purohit HJ (2011) Quenching the quorum sensing system: potential antibacterial drug targets. Crit Rev Microbiol 37:121-140.
12. Waters CM, Bassler BL (2005) Quorum sensing: cell-to-cell communication in bacteria. Annu Rev Cell Dev Biol 21:319-346.
13. Deep A, Chaudhary U, Gupta V (2011) Quorum sensing and bacterial pathogenicity: From molecules to disease. Journal of Laboratory Physicians 3:4-11.
14. Reading NC, Sperandio V (2006) Quorum sensing: the many languages of bacteria. FEMS Microbiol Lett 254:1-11.
15. Miller MB, Bassler BL (2001) Quorum sensing in bacteria. Annu Rev Microbiol 55:165.
16. Bottomley MJ, Muraglia E, Bazzo R, Carfì A (2007) Molecular insights into quorum sensing in the human pathogen Pseudomonas aeruginosa from the structure of the virulence regulator LasR bound to its autoinducer. J Biol Chem 282:13592-13600.
17. Rasmussen TB, Bjarnsholt T, Skindersoe ME, Hentzer M, Kristoffersen P, Köte M, Nielsen J, Eberl L, Givskov M (2005) Screening for Quorum-Sensing Inhibitors (QSI) by Use of a Novel Genetic System, the QSI Selector. J Bacteriol 187:1799-1814.
18. Zhang R, Pappas T, Brace JL, Miller PC, Oulmassov T, Molyneaux JM, Anderson JC, Bashkin JK, Winans SC, Joachimiak A (2002) Structure of a bacterial quorum-sensing transcription factor complexed with pheromone and DNA. Nature 417:971-974.
19. Sjöblom S, Brader G, Koch G, Palva ET (2006) Cooperation of two distinct ExpR regulators controls quorum sensing specificity and virulence in the plant pathogen Erwinia carotovora. Mol Microbiol 60:1474-1489.
20. Fuqua WC, Winans SC, Greenberg EP (1994) Quorum sensing in bacteria: the LuxR-LuxI family of cell density-responsive transcriptional regulators. J Bacteriol 176:269.
21. Swem LR, Swem DL, Wingreen NS, Bassler BL (2008) Deducing Receptor Signaling Parameters from In Vivo Analysis: LuxN/AI-1 Quorum Sensing in Vibrio harveyi. Cell 134:461-473.
22. Galloway WRJD, Hodgkinson JT, Bowden SD, Welch M, Spring DR (2010) Quorum Sensing in Gram-Negative Bacteria: Small-Molecule Modulation of AHL and AI-2 Quorum Sensing Pathways. Chem Rev 111:28-67.
23. Geske GD, O'Neill JC, Blackwell HE (2008) Expanding dialogues: from natural autoinducers to non-natural analogues that modulate quorum sensing in Gram-negative bacteria. Chem Soc Rev 37:1432-1447.

24. Galloway WRJD, Hodgkinson JT, Bowden S, Welch M, Spring DR (2012) Applications of small molecule activators and inhibitors of quorum sensing in Gram-negative bacteria. Trends Microbiol 20:449-458.
25. Skovstrup S, Quement L, Thordal S, Hansen T, Jakobsen TH, Harmsen M, Tolker-Nielsen T, Nielsen TE, Givskov M, Taboureau O (2013) Identification of LasR Ligands through a Virtual Screening Approach. Chem Med Chem 8:157-163.
26. Geske GD, Wezeman RJ, Siegel AP, Helen E (2005) Small molecule inhibitors of bacterial quorum sensing and biofilm formation. J Am Chem Soc 127:12762-12763.
27. Geske GD, O'Neill JC, Miller DM, Mattmann ME, Helen E (2007) Modulation of bacterial quorum sensing with synthetic ligands: systematic evaluation of N-acylated homoserine lactones in multiple species and new insights into their mechanisms of action. J Am Chem Soc 129:13613-13625.
28. Estephane J, Dauvergne J, Soulère L, Reverchon S, Queneau Y, Doutheau A (2008) N -Acyl-3-amino-5 H -furanone derivatives as new inhibitors of LuxR-dependent quorum sensing: Synthesis, biological evaluation and binding mode study. Bioorg Med Chem Lett 18:4321-4324.
29. McInnis CE, Blackwell HE (2011) Thiolactone modulators of quorum sensing revealed through library design and screening. Biorg Med Chem 19:4820-4828.
30. Nasser W, Reverchon S (2007) New insights into the regulatory mechanisms of the LuxR family of quorum sensing regulators. Anal Bioanal Chem 387:381-390.
31. Soulère L, Sabbah M, Fontaine F, Queneau Y, Doutheau A (2010) LuxR-dependent quorum sensing: Computer aided discovery of new inhibitors structurally unrelated to N-acylhomoserine lactones. Bioorg Med Chem Lett 20:4355-4358.
32. Sabbah M, Fontaine F, Grand L, Boukraa M, Efrit ML, Doutheau A, Soulère L, Queneau Y (2012) Synthesis and biological evaluation of new N-acyl-homoserine-lactone analogues, based on triazole and tetrazole scaffolds, acting as LuxR-dependent quorum sensing modulators. Biorg Med Chem 20:4727-4736.
33. Soulère L, Frezza M, Queneau Y, Doutheau A (2007) Exploring the active site of acyl homoserine lactones-dependent transcriptional regulators with bacterial quorum sensing modulators using molecular mechanics and docking studies. J Mol Graphics Model 26:581-590.
34. Soulère L, Guiliani N, Queneau Y, Jerez CA, Doutheau A (2008) Molecular insights into quorum sensing in Acidithiobacillus ferrooxidans bacteria via molecular modelling of the transcriptional regulator AfeR and of the binding mode of long-chain acyl homoserine lactones. J Mol Model 14:599-606.
35. Schrödinger Suite 2011 Protein Preparation Wizard; Epik version 2.2, Schrödinger, LLC, New York, NY, 2011; Impact version 5.7, Schrödinger, LLC, New York, NY, 2011; Prime version 3.0, Schrödinger, LLC, New York, NY, 2011.
36. Sadek MM, Serrya RA, Kafafy A-HN, Ahmed M, Wang F, Abouzid KAM (2013) Discovery of new HER2/EGFR dual kinase inhibitors based on the anilinoquinazoline scaffold as potential anti-cancer agents. J Enzyme Inhib Med Chem 1:1-8.
37. Ahmed M, Sadek MM, Serrya RA, Kafafy A-HN, Abouzid KA, Wang F (2013) Assessment of new anti-HER2 ligands using combined docking, QM/MM scoring and MD simulation. J Mol Graphics Model 40:91-98.
38. Prime version 3.0, Schrödinger, LLC, New York, NY, 2011.
39. Glide, version 5.7, Schrödinger, LLC, New York, NY, 2011.
40. Rocha GB, Freire RO, Simas AM, Stewart JJP (2006) RM1: A reparameterization of AM1 for H, C, N, O, P, S, F, Cl, Br, and I. J Comput Chem 27:1101-1111.
41. Jaguar, version 7.8, Schrödinger, LLC, New York, NY, 2011.
42. Friesner RA, Murphy RB, Repasky MP, Frye LL, Greenwood JR, Halgren TA, Sanschagrin PC, Mainz DT (2006) Extra Precision Glide:Docking and Scoring Incorporating a Model of Hydrophobic Enclosure for Protein−Ligand Complexes. J Med Chem 49:6177-6196.
43. D.A. Case, T.A. Darden, T.E. Cheatham, III, C.L. Simmerling, J. Wang, R.E. Duke, R. Luo, R.C. Walker, W. Zhang, K.M. Merz, B. Roberts, S. Hayik, A. Roitberg, G. Seabra, J. Swails, A.W. Goetz, I. Kolossváry, K.F. Wong, F. Paesani, J. Vanicek, R.M. Wolf, J. Liu, X. Wu, S.R. Brozell, T. Steinbrecher, H. Gohlke, Q. Cai, X. Ye, J. Wang, M.-J. Hsieh, G. Cui, D.R. Roe, D.H. Mathews, M.G. Seetin, R. Salomon-Ferrer, C. Sagui, V. Babin, T. Luchko, S. Gusarov, A. Kovalenko, and P.A. Kollman (2012), AMBER 12, University of California, San Francisco.
44. Duan Y, Wu C, Chowdhury S, Lee MC, Xiong G, Zhang W, Yang R, Cieplak P, Luo R, Lee T, Caldwell J, Wang J, Kollman P (2003) A point-charge force field for molecular mechanics simulations of proteins based on condensed-phase quantum mechanical calculations. J Comput Chem 24:1999-2012.
45. M.J.Frisch et al. (2009) Gaussian09, Rev A.02, Gaussian Inc,Wallingford, CT.
46. Bayly CI, Cieplak P, Cornell W, Kollman PA (1993) A well-behaved electrostatic potential based method using charge restraints for deriving atomic charges: the RESP model. J Phys Chem 97:10269-10280.
47. Wang J, Wolf RM, Caldwell JW, Kollman PA, Case DA (2004) Development and testing of a general amber force field. J Comput Chem 25:1157-1174.
48. Jorgensen WL, Chandrasekhar J, Madura JD, Impey RW, Klein ML (1983) Comparison of simple potential functions for simulating liquid water. J Chem Phys 79:926-935.
49. Ryckaert J-P, Ciccotti G, Berendsen HJC (1977) Numerical integration of the cartesian equations of motion of a system with constraints: molecular dynamics of n-alkanes. J Comput Phys 23:327-341.
50. Berendsen HJC, Postma JPM, van Gunsteren WF, DiNola A, Haak JR (1984) Molecular dynamics with coupling to an external bath. J Chem Phys 81:3684-3690.
51. Darden T, York D, Pedersen L (1993) Particle mesh Ewald: An N.log(N) method for Ewald sums in large systems. J Chem Phys 98:10089-10092.
52. Miller BR, McGee TD, Swails JM, Homeyer N, Gohlke H, Roitberg AE (2012) MMPBSA.py: An Efficient Program for End-State Free Energy Calculations. Journal of Chemical Theory and Computation 8:3314-3321.
53. Schwobel J, Ebert R, Kuhne R, Schuurmann G (2009) Prediction of the Intrinsic Hydrogen Bond Acceptor Strength of Chemical Substances from Molecular Structure. J Phys Chem A 113:10104-10112.
54. Ho BK, Agard DA (2009) Probing the flexibility of large conformational changes in protein structures through local perturbations. PLoS Comp Biol 5:1000343.
55. Stauff DL, Bassler BL (2011) Quorum Sensing in Chromobacterium violaceum: DNA Recognition and Gene Regulation by the CviR Receptor. J Bacteriol 193:3871-3878.
56. Voter AF (1997) A method for accelerating the molecular dynamics simulation of infrequent events. The Journal of Chemical Physics 106:4665-4677.
57. Lefèvre F, Rémy M-H, Masson J-M (1997) Alanine-stretch scanning mutagenesis: a simple and efficient method to probe protein structure and function. Nucleic Acids Res 25:447-448.
58. Morrison KL, Weiss GA (2001) Combinatorial alanine-scanning. Curr Opin Chem Biol 5:302-307.

**Table 1:** Selected binding energy scores together with the *in vitro* measured $IC_{50}$ values for the inhibitors under study ranked according to the XP Gscore.

| Inhibitors | $IC_{50}$ (μM)[a] | XP GScore | XP LipophilicEvdW | XP Electrostatic | Glide Emodel | Prime MMGBSA DG bind | Prime MMGBSA DG bind vdW |
|---|---|---|---|---|---|---|---|
| L14 | | -10.28 | -4.90 | -1.16 | -91.15 | -99.19 | -49.53 |
| L15 | | -10.25 | -5.12 | -1.19 | -96.06 | -102.41 | -47.95 |
| L12 | | -10.02 | -4.19 | -1.19 | -89.99 | -100.53 | -46.46 |
| L11 | | -9.71 | -4.66 | -1.19 | -93.41 | -116.59 | -51.84 |
| L4 | | -9.50 | -4.60 | -1.20 | -94.81 | -98.70 | -45.54 |
| L3 | 0.38 | -9.49 | -4.63 | -1.20 | -90.40 | -105.54 | -46.19 |
| TL12 | 0.63 | -9.43 | -4.14 | -1.07 | -80.56 | -104.45 | -44.09 |
| L9 | | -9.40 | -4.50 | -1.18 | -96.34 | -107.02 | -48.73 |
| L13 | | -9.40 | -4.40 | -1.21 | -70.66 | -101.06 | -47.33 |
| L5 | | -9.38 | -4.64 | -1.17 | -95.11 | -106.60 | -48.48 |
| L8 | | -9.36 | -4.48 | -1.17 | -92.92 | -102.88 | -44.60 |
| TL15 | 5.00 | -9.34 | -4.76 | -1.09 | -89.31 | -102.13 | -46.57 |
| L2 | | -9.34 | -4.29 | -1.18 | -91.56 | -94.47 | -43.61 |
| TL14 | 4.00 | -9.32 | -4.50 | -1.05 | -83.97 | -101.76 | -46.95 |
| TL1 | | -9.11 | -4.27 | -1.18 | -89.77 | -92.72 | -43.31 |
| TL11 | 1.40 | -9.08 | -4.49 | -1.09 | -86.47 | -118.22 | -51.14 |
| L10 | | -8.76 | -4.31 | -1.10 | -72.97 | -99.16 | -45.17 |
| TL3 | 1.10 | -8.61 | -4.19 | -1.05 | -79.07 | -104.49 | -45.00 |
| TL4 | 1.80 | -8.60 | -4.25 | -1.07 | -82.58 | -101.97 | -45.27 |
| TL2 | 2.10 | -8.58 | -3.86 | -1.06 | -81.62 | -100.98 | -39.87 |
| TL5 | 1.40 | -8.53 | -4.27 | -1.07 | -83.29 | -102.32 | -45.68 |
| TL8 | 2.10 | -8.51 | -3.97 | -0.98 | -81.19 | -115.31 | -48.53 |
| TL9 | 2.70 | -8.36 | -4.11 | -1.04 | -80.41 | -109.05 | -41.12 |
| TL13 | 3.40 | -8.20 | -3.75 | -1.02 | -69.40 | -95.52 | -40.07 |
| TL1 | 2.60 | -8.18 | -3.91 | -1.07 | -79.01 | -96.20 | -42.11 |
| L6 | | -8.05 | -3.92 | -1.13 | -77.15 | -90.41 | -42.70 |
| L7 | | -7.81 | -4.26 | -1.09 | -74.02 | -99.39 | -44.11 |
| L16 | | -7.78 | -4.66 | -1.18 | -76.10 | -101.75 | -48.56 |
| TL10 | 37.00 | -7.67 | -3.92 | -0.94 | -70.76 | -97.99 | -37.71 |
| TL6 | 2.50 | -7.29 | -4.12 | -1.14 | -75.55 | -112.28 | -47.43 |
| TL7 | 11.00 | -6.95 | -4.05 | -0.88 | -65.21 | -98.33 | -38.84 |
| TL16 | 3.90 | -5.69 | -3.95 | -0.84 | -77.05 | -109.95 | -49.65 |
| $r_{(Pearson)}$[b] | | 0.30 | 0.25 | 0.44 | 0.49 | 0.34 | 0.52 |

[a]See Ref. [8].

[b] Pearson correlation coefficient between the *in silico* calculated docking scores and the *in vitro* measured $IC_{50}$ values.

**Table 2:** Total and decomposed binding energies of the MD studied complexes (kcal·mol$^{-1}$) together with the experimental $IC_{50}$ values.

| **Contribution** | **Inhibitor** | |
|---|---|---|
| | **L3** | **TL3** |
| **$\Delta E_{ele}$** | **-33.54** | **-28.37** |
| **$\Delta E_{vdW}$** | **-45.31** | **-46.56** |
| **$\Delta E_{binding\text{-}gas}$[a]** | **-78.85** | **-74.93** |
| **ΔGbinding (AMBER-PBSA)[b]** | **-63.54** | **-59.11** |
| **ΔGbinding (AMBER-GBSA)[b]** | **-48.65** | **-47.60** |

[a] $\Delta \boldsymbol{E}_{\mathbf{binding\text{-}gas}} = \Delta \boldsymbol{E}_{\mathbf{ele}} + \Delta \boldsymbol{E}_{\mathbf{vdw}}$

[b] $\Delta \boldsymbol{G}_{\mathbf{binding}} = \Delta \boldsymbol{E}_{\mathbf{ele}} + \Delta \boldsymbol{E}_{\mathbf{vdw}} + \Delta \boldsymbol{G}_{\mathbf{solvation}}$

**Figure captions**

**Figure 1**: Two dimensional (2D) structures of the lactone (X=O) and thiolactone (X=S) inhibitors under study.

**Figure 2**: Three dimensional (3D) ribbon representation for the CviR monomer. Solid surface in the LBD represent the exact binding location of the inhibitors.

**Figure 3**: 2D ligand interaction diagrams for (a) a potential CviR antagonist (PDB code: 3QP5) and (b) a potential CviR agonist (PDB code: 3QP1).

**Figure 4**: 2D and 3D ligand interactions for some selected inhibitors under study in the CviR binding site (a) L12, (b)TL12 and (c) L14.

**Figure 5**: RMSD plots for the protein backbone Cα atoms during the 30 ns production simulation for (a) dimer complexes and (b) monomer complexes.

**Figure 6**: Per-residue heavy atoms RMSF plots during the 30 ns production simulation for (a) dimer complexes and (b) monomer complexes.

**Figure 7**: Selected H-bonds distances monitored during the 30 ns production simulations for (a, c) dimer complexes and (b, d) monomer complexes.

**Figure 8**: 3D representation for (a) **L3** and (b) **TL3** with the surrounding amino acid residues. These two snapshots were obtained at the end of the 30 ns production simulations.

**Figure 9**: Per-residue binding energy decompositions calculated using the MMPBSA module of AMBER, (a) per-residue contribution to the vdW interaction term and (b) per-residue contribution to the electrostatic interaction term.

**Figure 1**: Two dimensional (2D) structures of the lactone (X=O) and thiolactone (X=S) inhibitors under study.

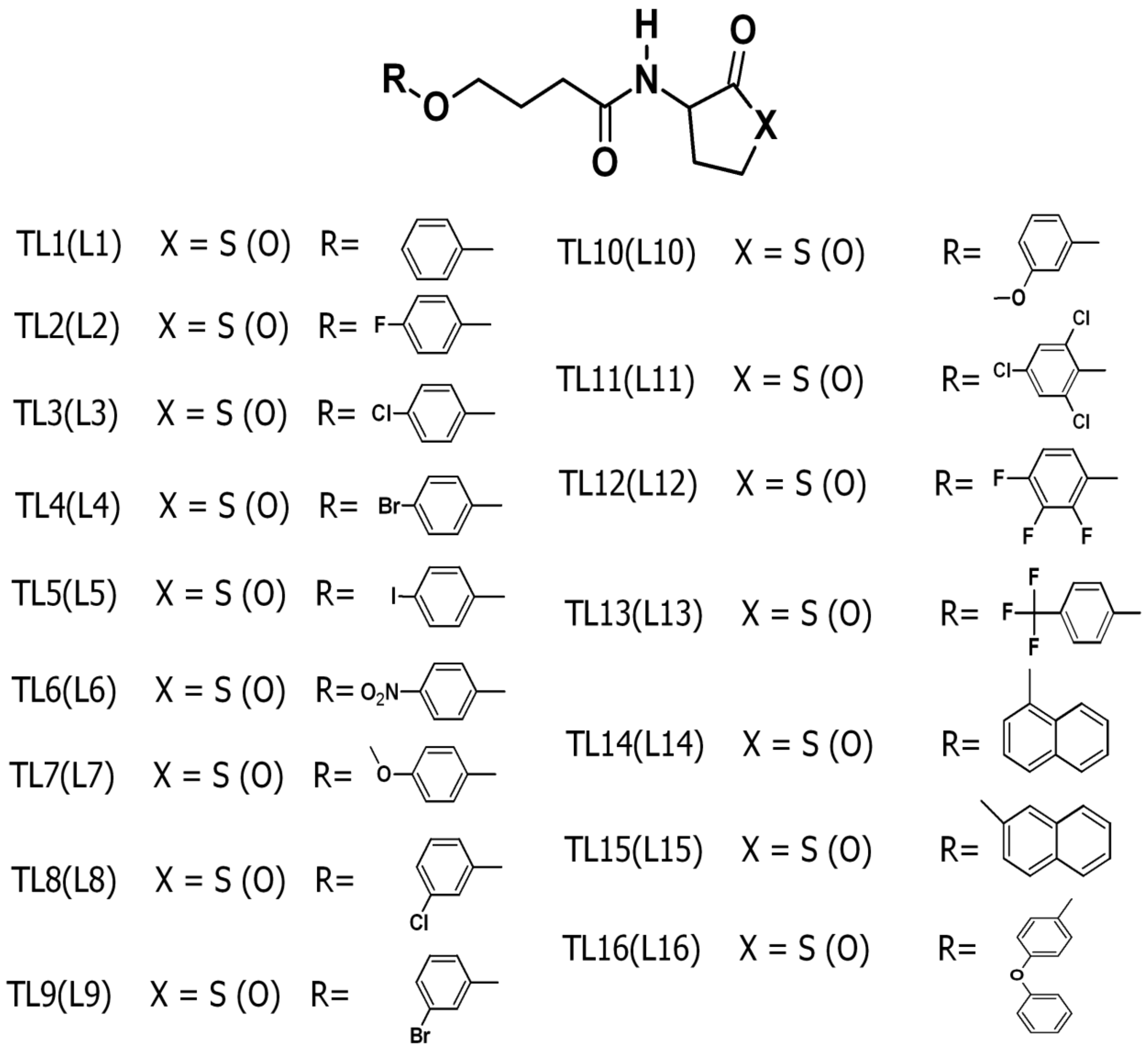

**Figure 2**: Three dimensional (3D) ribbon representation for the CviR monomer. Solid surface in the LBD represent the exact binding location of the inhibitors.

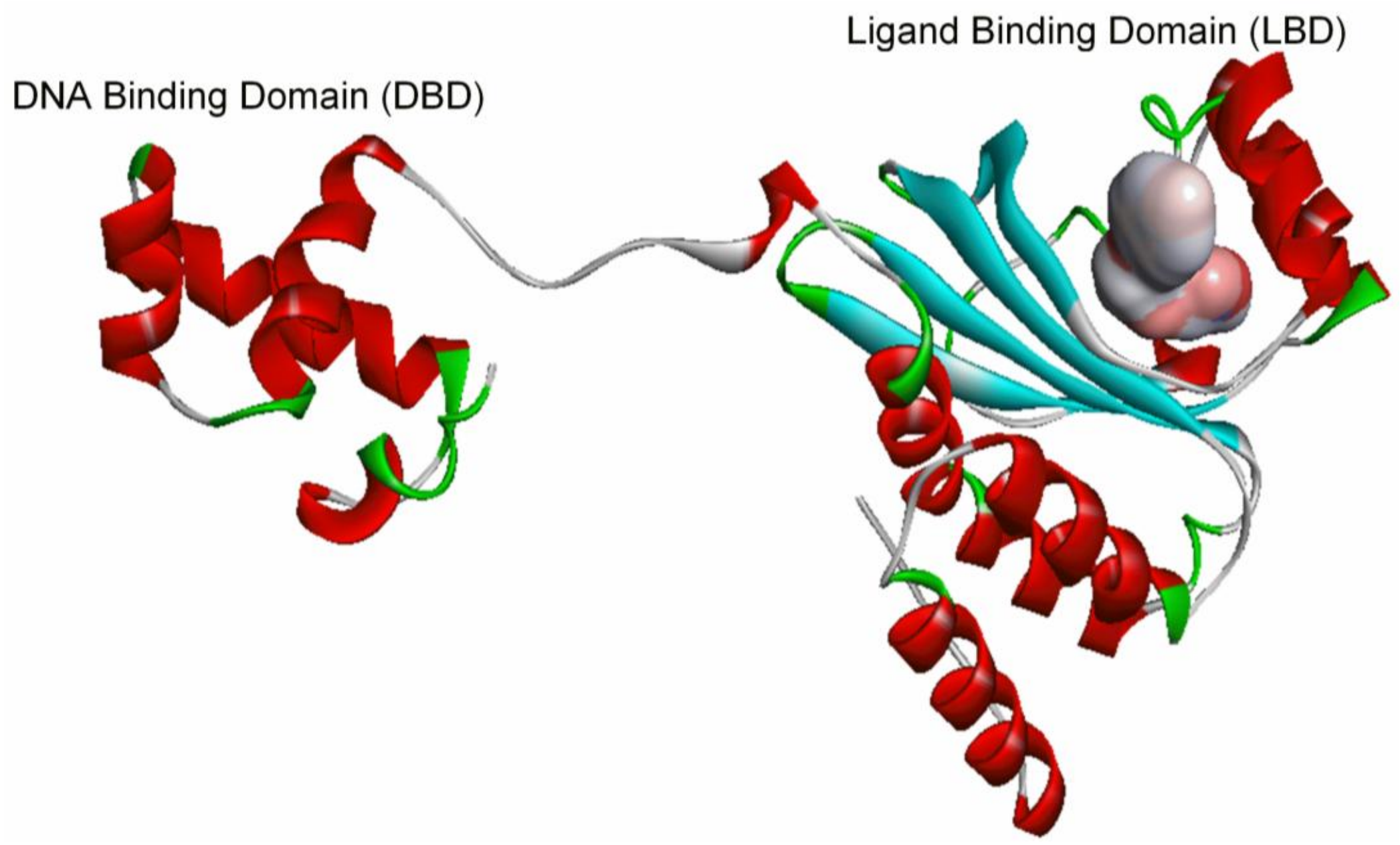

**Figure 3**: 2D ligand interaction diagrams for (a) a potential CviR antagonist (PDB code: 3QP5) and (b) a potential CviR agonist (PDB code: 3QP1).

(a)

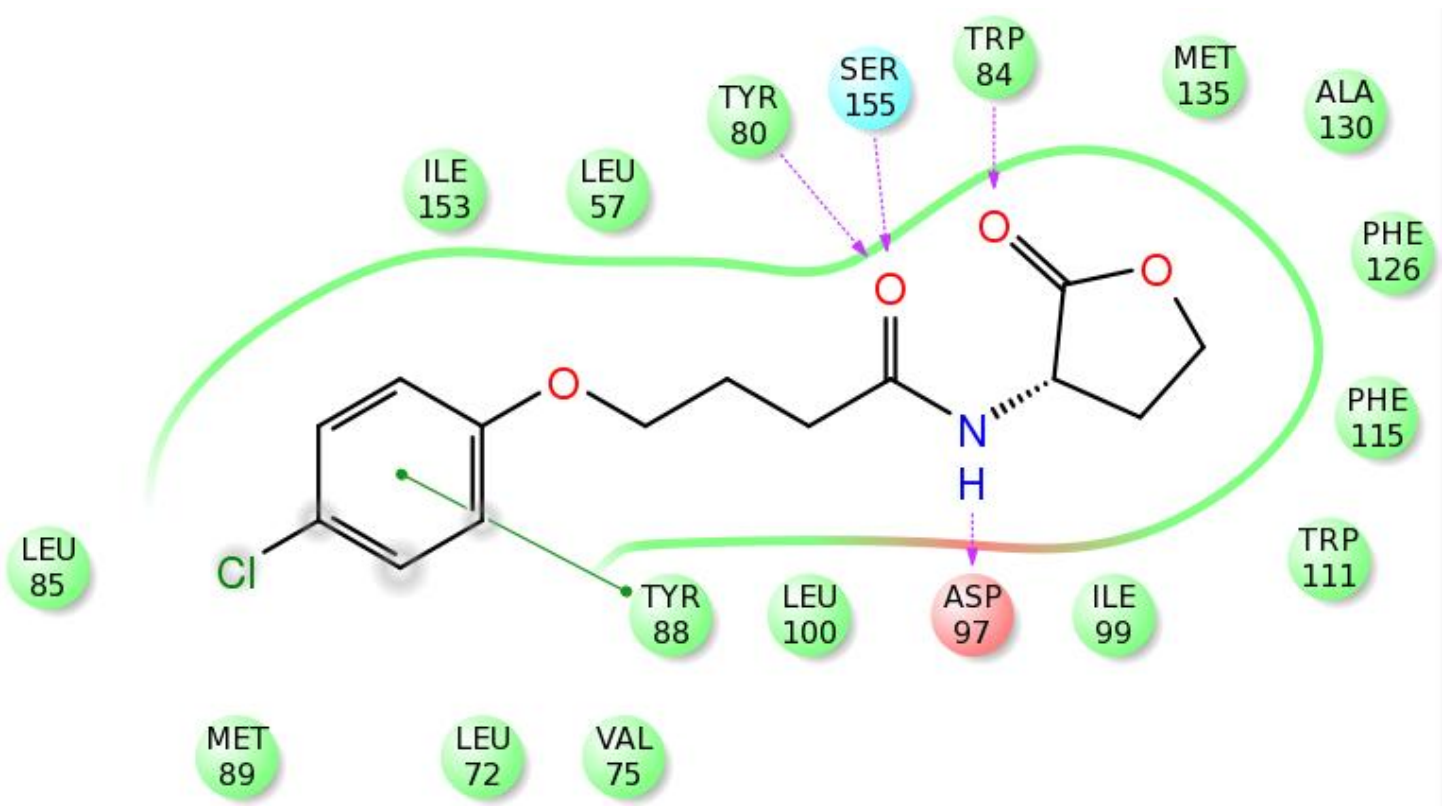

(b)

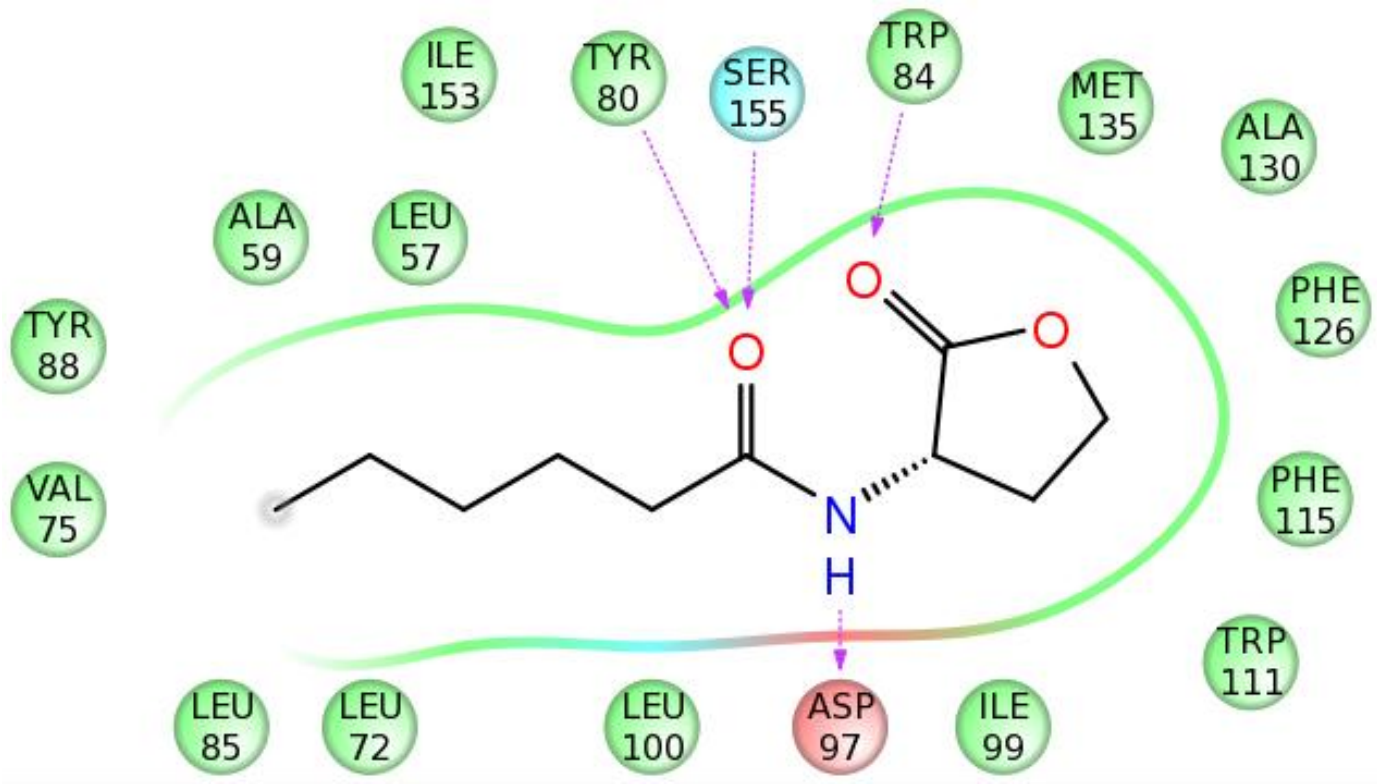

**Figure 4**: 2D and 3D ligand interactions for some selected inhibitors under study in the CviR binding site (a) L12, (b)TL12 and (c) L14.

(a)

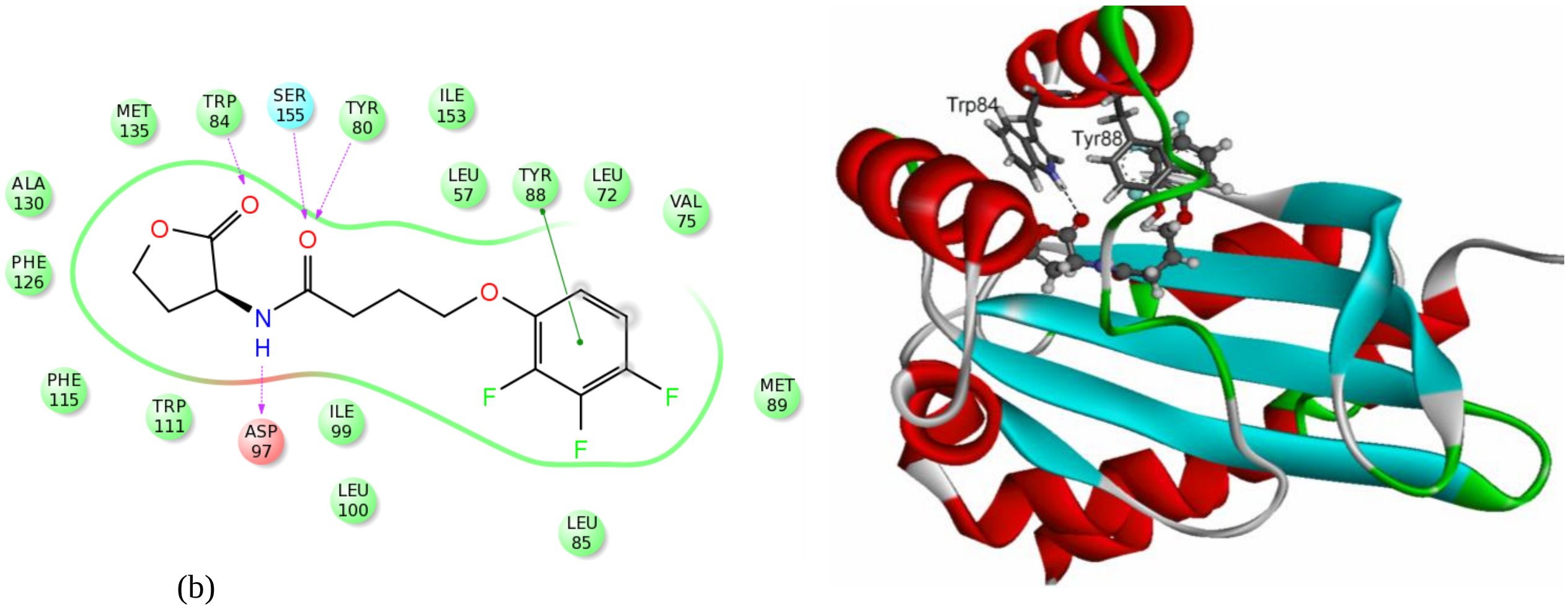

(b)

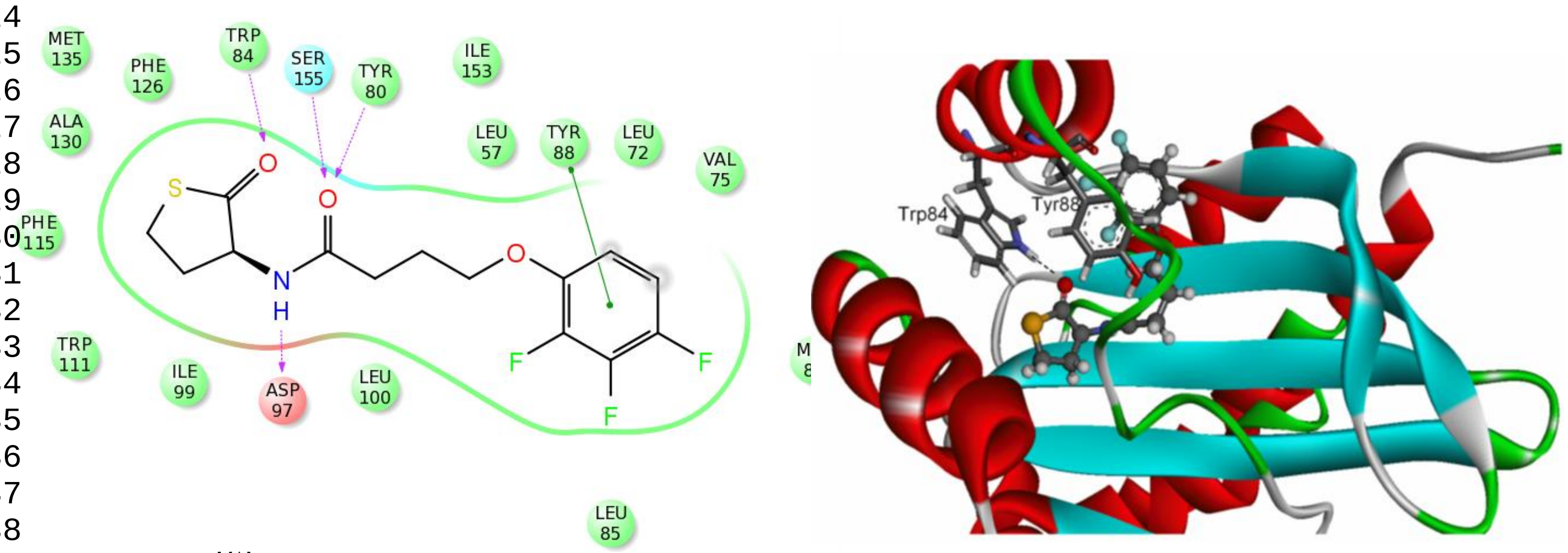

(c)

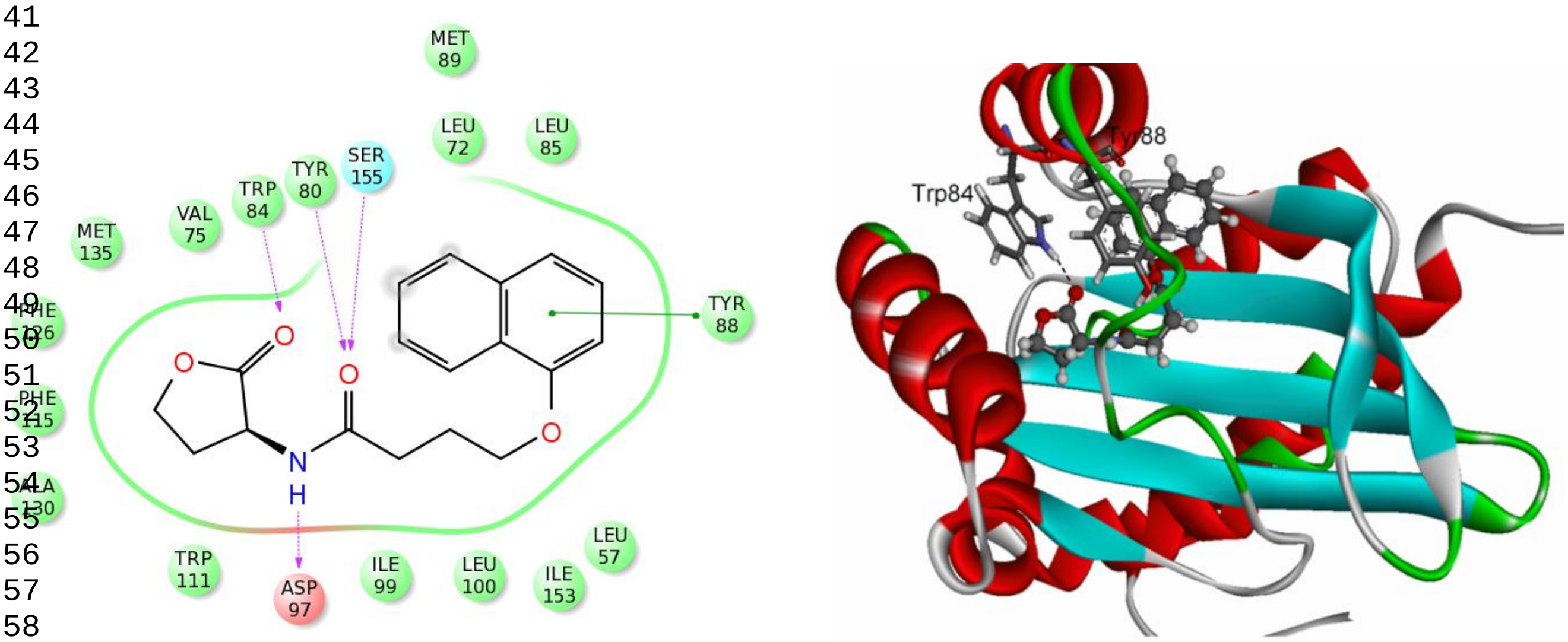

**Figure 5**: RMSD plots for the protein backbone Cα atoms during the 30 ns production simulation for (a) dimer complexes and (b) monomer complexes.

(a)

(b)

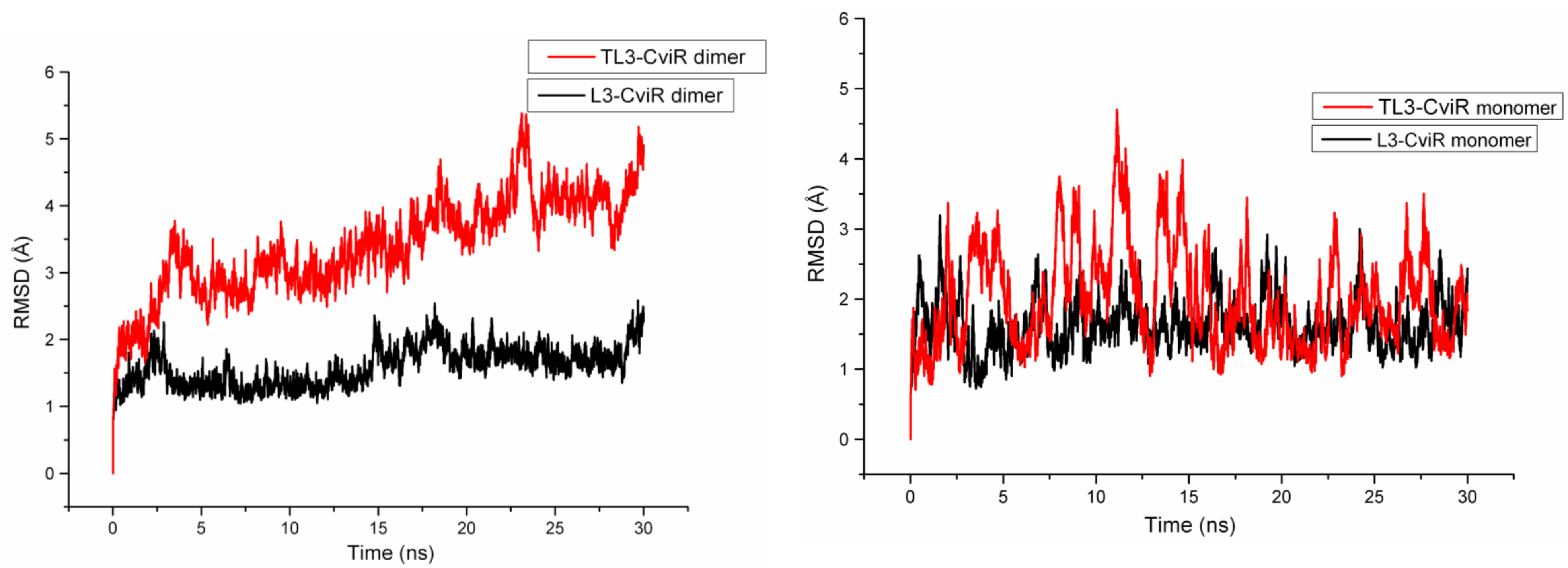

**Figure 6**: Per-residue heavy atoms RMSF plots during the 30 ns production simulation for (a) dimer complexes and (b) monomer complexes.
.

(a) (b)

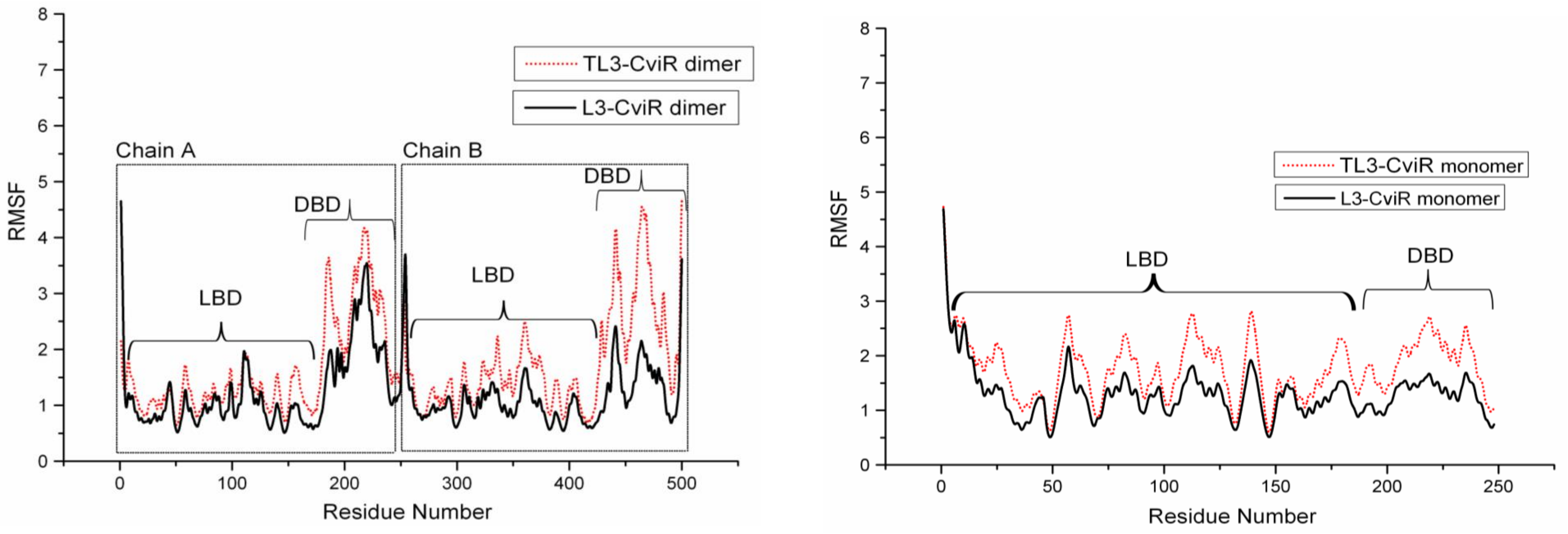

**Figure 7**: Selected H-bonds distances monitored during the 30 ns production simulations for (a, c) dimer complexes and (b, d) monomer complexes.

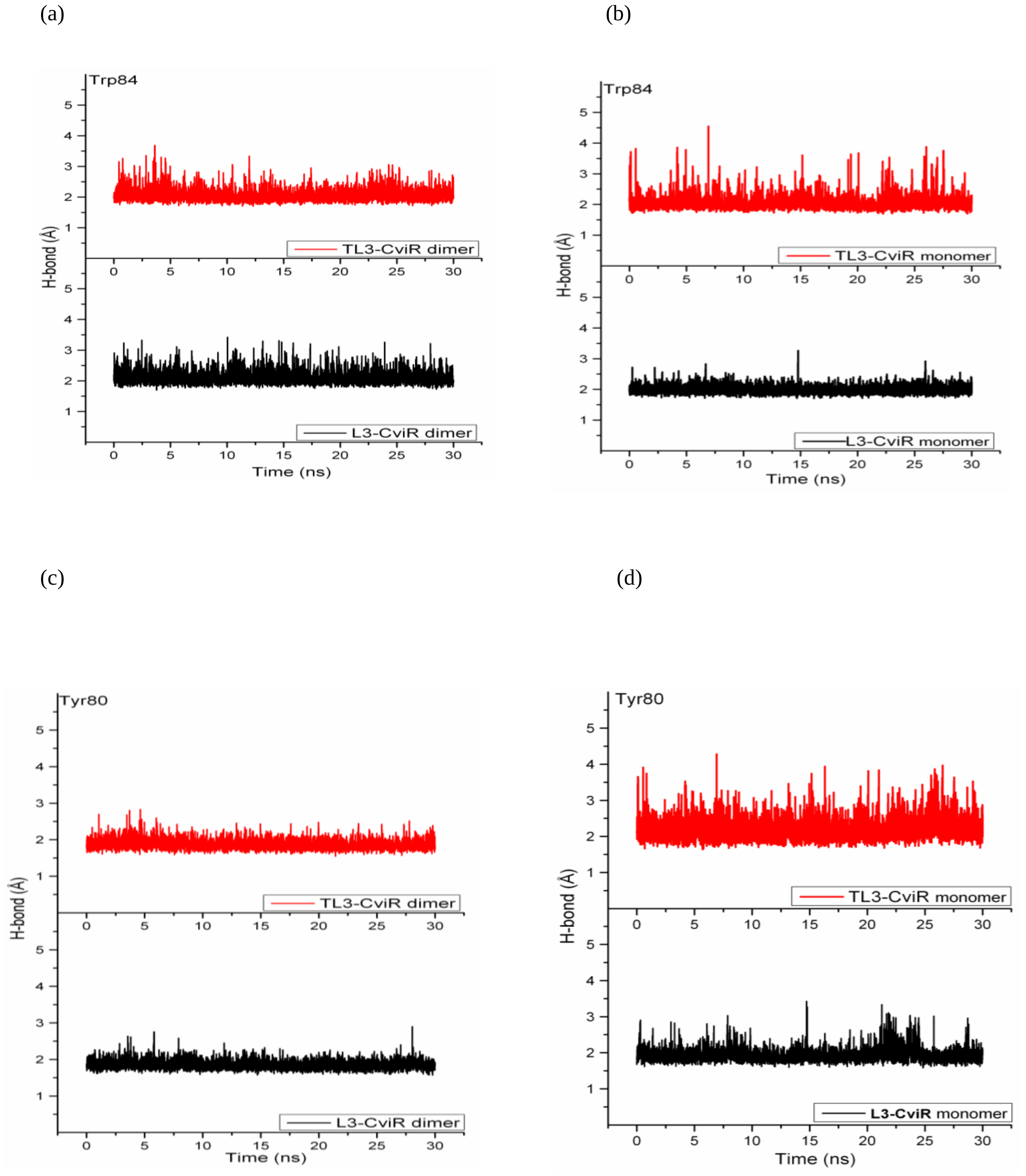

**Figure 8**: 3D representation for (a) **L3** and (b) **TL3** with the surrounding amino acid residues. These two snapshots were obtained at the end of the 30 ns production simulations.

(a)

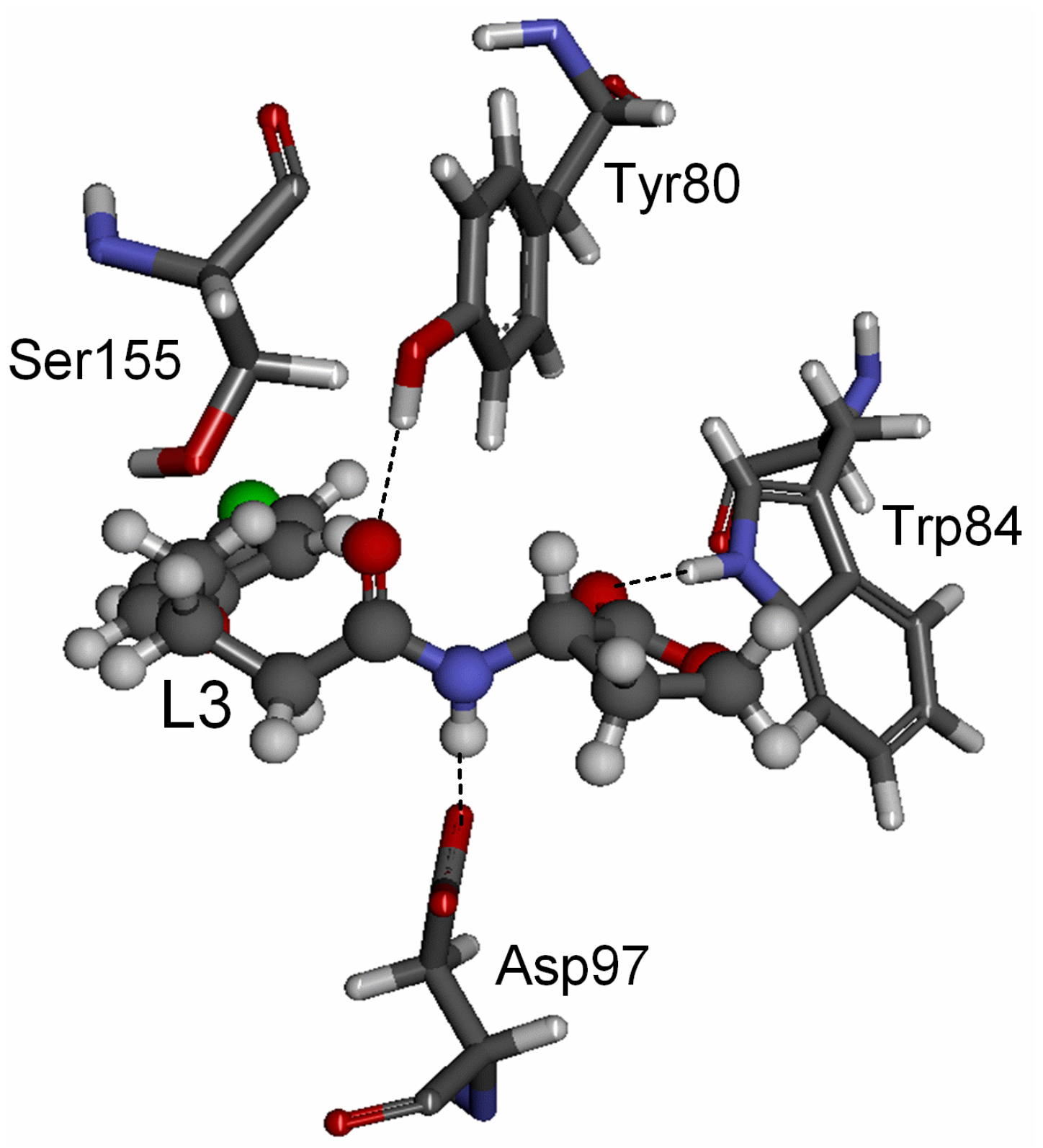

(b)

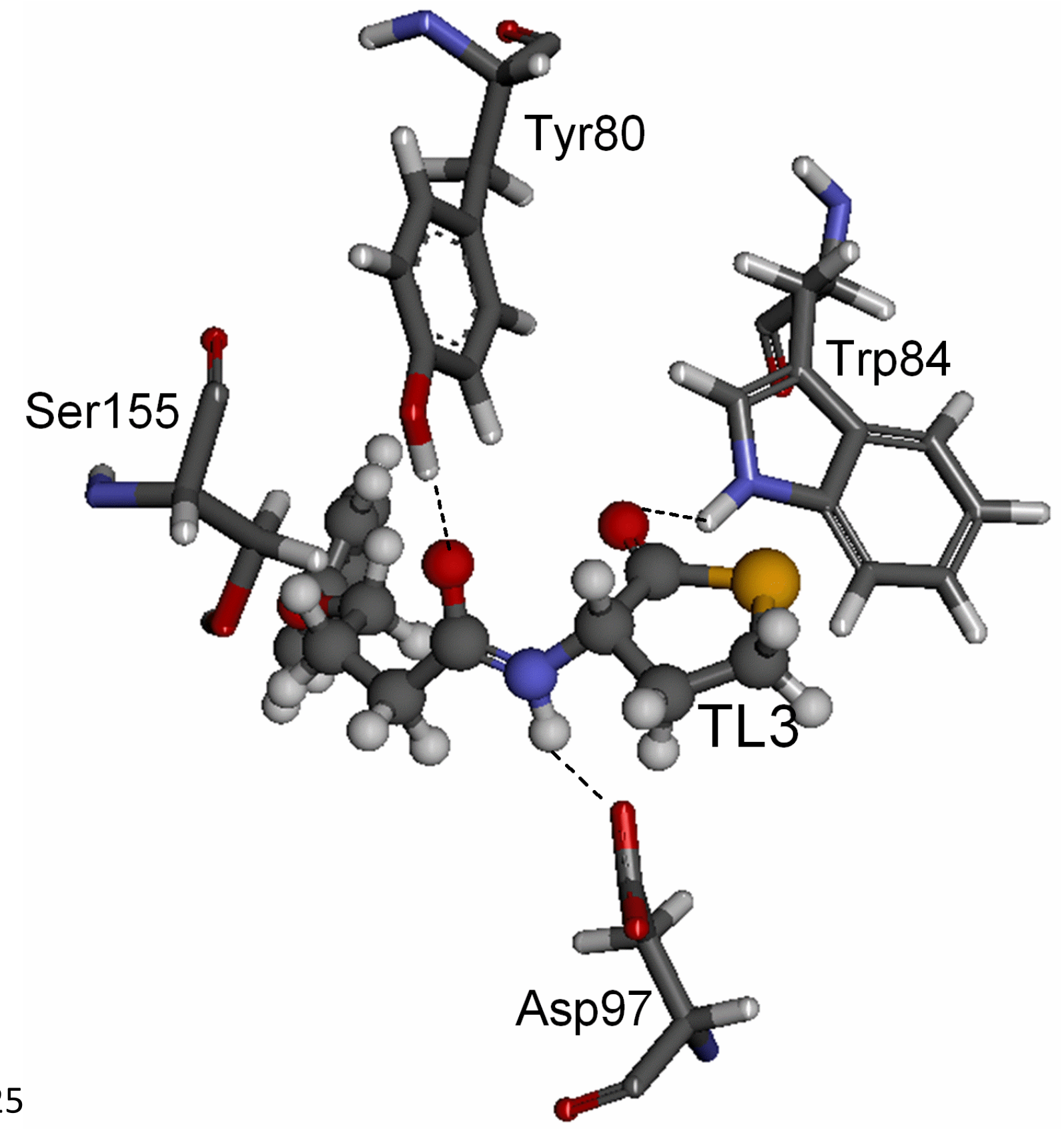

**Figure 9**: Per-residue binding energy decompositions calculated using the MMPBSA module of AMBER, (a) per-residue contribution to the vdW interaction term and (b) per-residue contribution to the electrostatic interaction term.

(a)

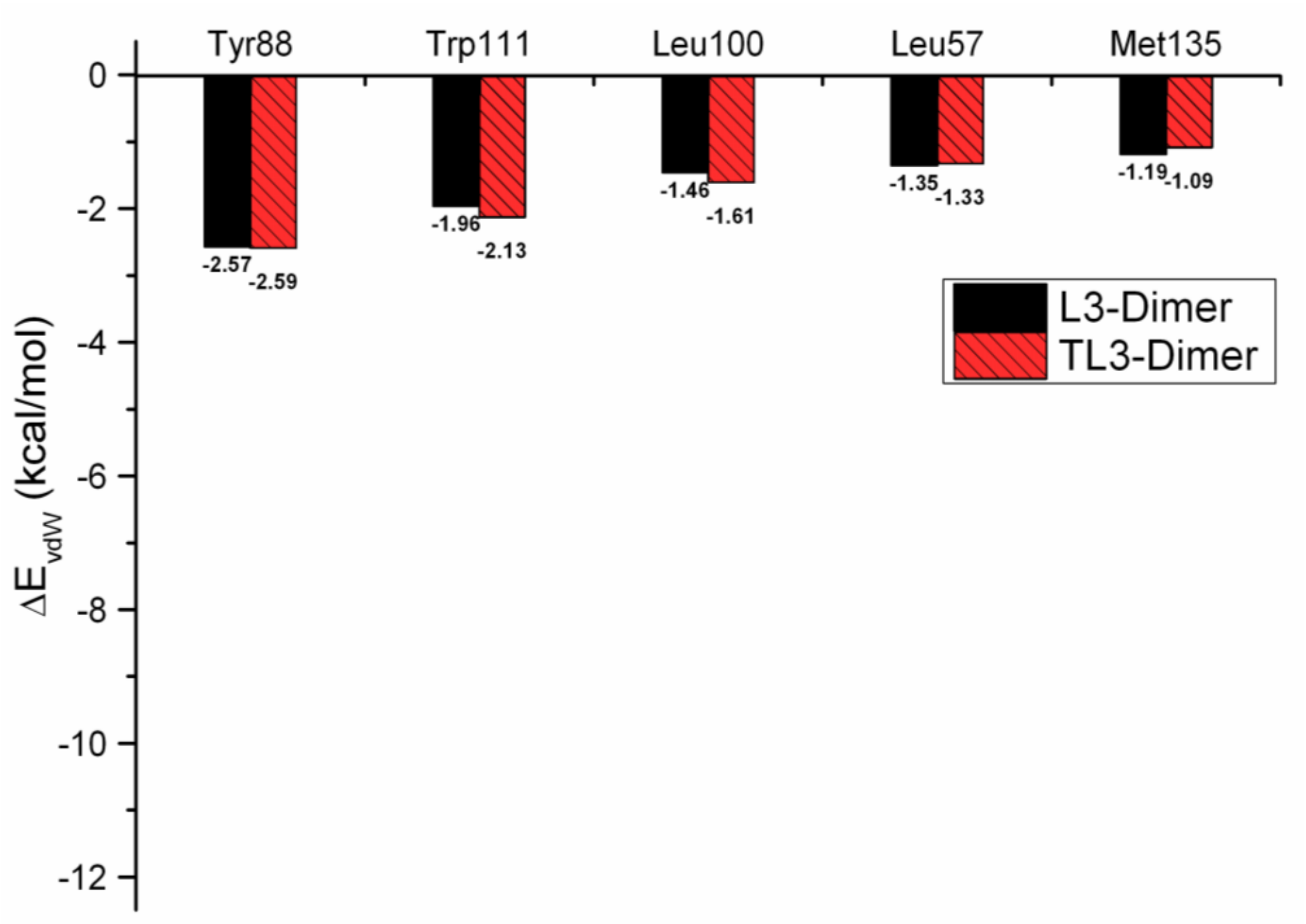

(b)

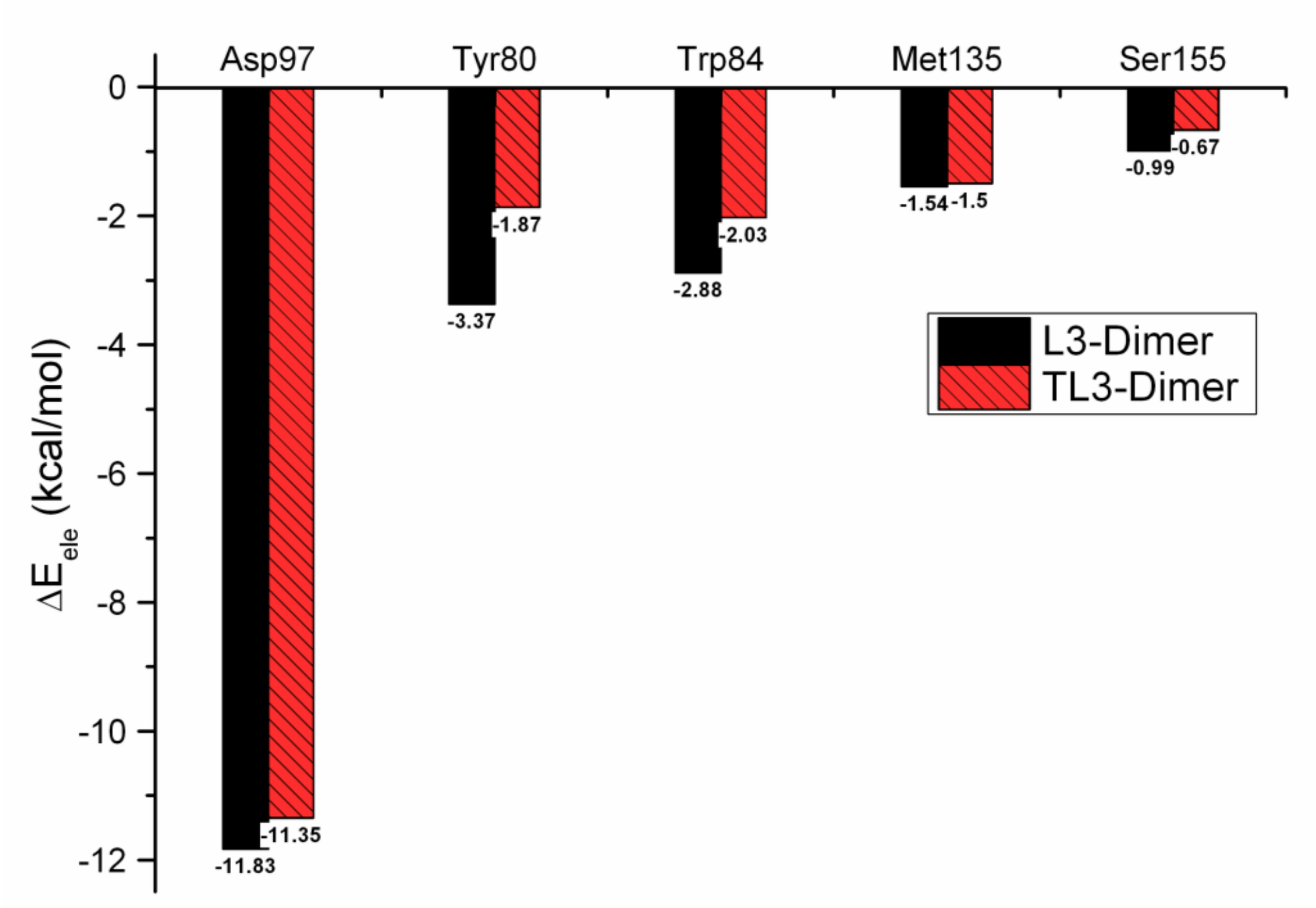